\documentclass{emulateapj}
\usepackage{apjfonts}
\newcommand\beq{\begin{equation}}
\newcommand\eeq{\end{equation}}

\newcommand\kpc{$h^{-1}$\,{\rm \,kpc}}

\newcommand\Mpc{$h^{-1}$\,{\rm \,Mpc}}
\newcommand\Msun{$h^{-1}\,M_\odot$}

\shorttitle{Mass Distribution of Dark Matter Halos}
\shortauthors{Lin et al.}

\begin{document}

\title{The Influence of Baryons on the Mass Distribution of Dark Matter Halos}

\author{W.P. Lin \altaffilmark{1,4}, Y.P. Jing\altaffilmark{1,4}, S. Mao\altaffilmark{2}, L. Gao\altaffilmark{3}, I.G. McCarthy\altaffilmark{3}}
\altaffiltext{1}{Shanghai Astronomical Observatory, Nandan Road 80,
Shanghai 200030, China}
\altaffiltext{2}{University of Manchester, Jodrell Bank Observatory, Macclesfield, Cheshire SK11 9DL, U.K.}
\altaffiltext{3}{Institute for Computational Cosmology, Physics Department, Durham, DH1 3LE, U.K.}
\altaffiltext{4}{Joint Institute for Galaxy and Cosmology (JOINGC) of SHAO and USTC}
\email{linwp,ypjing@shao.ac.cn,\\smao@jb.man.ac.uk,\\liang.gao,i.g.mccarthy@durham.ac.uk}

\begin{abstract}
Using a set of high-resolution $N$-body/SPH cosmological simulations
with identical initial conditions but run with different
numerical setups, we investigate the influence of baryonic matter on
the mass distribution of dark halos when radiative
cooling is {\it not} included. We compare the
concentration parameters of about $400$ massive halos with
virial mass from $10^{13}$ to $7.1  \times
10^{14}$ \Msun. We find that the concentration parameters for the
total mass and dark matter distributions in nonradiative simulations
are on average larger by $\sim 3\%$ and 10\% than those in a pure dark matter
simulation. 
Our results indicate that the total mass density profile is little affected by a hot gas
component in the simulations.
After carefully excluding the effects of
resolutions and spurious two-body heating between dark matter (DM) and gas 
particles we conclude that the increase of the DM
concentration parameters is due to interactions between baryons and DM.
We demonstrate this with the aid of idealized simulations of two-body mergers.
The results of individual halos simulated with different
mass resolutions show that the gas profiles of densities, temperature and
entropy are subjects of mass resolution of SPH particles. In particular,
we find that in the inner parts of halos, as the
SPH resolution increases the gas density becomes higher but both the
entropy and temperature decrease.

\end{abstract}


\section{INTRODUCTION}
\label{introduction}
One potential problem of the cold dark matter (CDM) cosmology
is the mass distributions in the central region of clusters of galaxies.  
There is yet no consensus emerging from the observations.
In general, simulations produce steep inner density profiles for
DM halos \citep{NFW97, Moore98, JS00}, 
while observations of some clusters seem to prefer flat, core-like
profiles \citep[e.g.,][]{Tyson98}, and others prefer cusped profiles \citep[e.g.,][]{Lewis03,Buote04,Pointecouteau05}. 
\cite{Sand04} show that the central DM profiles of clusters of galaxies
have a slope of about $-0.5$, which is substantially flatter than the inner
slope ($-1$) of the Navarro-Frenk-White (NFW) profile as found in CDM simulations. However, other
authors argued that their interpretations are compromised by the
assumption of sphericity \citep[e.g.,][]{Bartelmann04,Meneghetti05, Dalal03}. 
Recent observations of strong gravitational lensing
have shown that some clusters of galaxies have very high mass
concentration parameters \citep[e.g.,][]{Broadhurst05} compared with
the average value obtained in the DM simulations.
The  number of giant arcs predicted by $N$-body simulations
may also be somewhat too low to be compatible with observations
(\citealt{Li05}; but see \citealt{Dalal04}; \citealt{Hen06};
\citealt{Horesh05}).
It is unclear whether these ''discrepancies'' are serious.
Hence, the systematics in theoretical predictions need to be 
explored, and the effect of baryons is one of these.

Baryons can cool via dissipative processes while DM
is collisionless, and so their density evolutions may be quite
different. Although the baryonic matter is only a modest fraction
of the total mass (less than or equal to the universal baryon
fraction of $\sim$ 16\%, as revealed by the {\em Wilkinson Microwave Anistropy Probe} [WMAP]; Spergel et al. 2003, 2006),
it can nevertheless play a significant role in reshaping the density
profiles of the DM halos.  For example, the baryons that condense toward
the center of DM halos can compress DM by adiabatic contraction
\citep{Blumenthal86,Mo98}. This effect was confirmed in 
simulations with gas cooling 
where the DM density profiles appear steeper than those in the pure DM simulations
\citep[e.g.,][]{Pearce00,Gnedin04}.
Even for nonradiative gas simulations, several studies \citep{Pearce94,
Navarro95, Rasia04, Jing06} suggest that the mass density profiles of DM halos
become steeper.
Some analytic studies \citep[e.g.,][]{Zhan04} often adopt DM
profiles (such as the NFW profile) straight
from DM-only simulations and do not
take into account the modification of the DM profiles by baryons.
In this paper, we explore the effects of baryons using high-resolution simulations 
with nonradiative gas.
In this case, the interaction between the hot gas and DM is a well-defined
problem, and hence our study sets a benchmark for the effects of
baryons on the matter distribution when important (but more uncertain) 
star formation and feedback processes are incorporated.
To do this, we perform a set of $N$-body/SPH simulations with different mass 
and force resolutions and compare the results with those in a pure DM 
simulation. 
We will examine effects of resolutions and spurious two-body 
heatings between the gas and DM particles (\citealt{Steinmetz97}).
The layout of this paper is as follows. In \S2 we give an
introduction of our simulations and methods; in \S3 we present
our main results. We examine the hypothesis of energy transfer from 
DM to gas in \S4, and finish with a summary and discussion in \S5.

\section{THE SIMULATIONS AND METHODS}

We use the massively parallel GADGET2 code \citep{Springel01, Springel05} to simulate
structure formation. The code can follow a collisionless fluid with the
$N$-body method, and an ideal gas by means of smoothed particle
hydrodynamics (SPH). The GADGET2 implementation of SPH conserves energy
and entropy (\citealt{Springel02}). The simulations were performed in
a concordance cosmological model with the
following parameters $(\Omega_{m},\Omega_{\Lambda},\Omega_{b},\sigma_8,n,h)$
$=(0.268,0.732,0.044,0.85,1,0.71)$. Five simulations  were
run with the same initial condition in a cubic box of sidelength $100$ \Mpc\,
(see Table 1 for details). A pure DM (PDM) simulation is used
as a comparison for the simulations that include baryons. In the four
simulations with gas, no radiative cooling is considered and the
gas is treated as an ideal gas with an adiabatic index $\gamma=5/3$.

\begin{table}
\caption{Simulation parameters}
\begin{tabular}{lccccc}
\hline
Run Name & $N_{\rm DM}$ & $N_{\rm gas}$ & $m_{\rm DM}$  & $m_{\rm gas}$ & Softening Length \\
    &          &           &   \Msun        &  \Msun         &  \kpc \\
\hline
PDM     & $512^3$  & -       & $5.5\times 10^8$  &  -  &  4.5  \\
A1  & $256^3$  & $256^3$ & $3.7\times 10^9$  & $7.4\times 10^8$   &  4.5  \cr
A2  & $256^3$  & $256^3$ & $3.7\times 10^9$  & $7.4\times 10^8$   &  9.0  \cr
A3  & $512^3$  & $256^3$ & $4.6\times 10^8$  & $7.4\times 10^8$   &  4.5  \cr
A4  & $512^3$  & $512^3$ & $4.6\times 10^8$  & $9.2\times 10^7$   &  4.5  \cr
\hline
\end{tabular}
\label{table1}
\end{table}

The highest resolution simulation we performed is A4,
which includes an equal number ($512^3$) of gas and DM particles.
The simulation A1 has the lowest resolution, but
it has the smallest softening length relative to the
mean interparticle separation, about $\sim 1/80$ ($\sim 1/40$ for other simulations). 
A1 is used to check the effects of spurious large accelerations
in close approaches between particles (the real mass distribution is
inherently smooth and unable to generate large accelerations).
In run A2, we double the softening length from A1 to check its effects.
We also ran a simulation A3,
in which gas and DM particles have comparable mass to 
study spurious two-body heating effects, which depend on the number of
DM particles used (\citealt{Steinmetz97}).
Two-body heating occurs when DM particles collide with gas particles, 
lose their kinetic energy to the gas
components, and, as a result become more concentrated. This effect is cumulative
in a cosmological simulation and may be particularly serious in
simulations with low mass resolutions or for halos with a small
number of particles. Thus it has been suggested that it is wise to use
more DM particles than gas particles for cosmological simulations
(\citealt{Steinmetz97}); the A3 run fulfills this requirement.
Table 1 lists the number and mass of particles, and the softening length used in the simulations.

The simulations discussed above are all evolved from redshift $z=120$ to the present
epoch. Because the initial conditions are 
identical, one can match each well resolved halo on an one-to-one basis
in different runs. Thus it is straightforward
to select the massive mass halos at $z=0$ and compare their radially averaged density
profiles in different simulations.
The NFW form (Navarro et al. 1997) is used to fit the halo density
profiles. The NFW profile is given by $\rho(x)={\Delta \bar{\rho}
f(c)}/c x/(1+c x)^2$, where $\Delta=200$ is the overdensity,
$\bar{\rho}$ is the mean universe mass density, $x=r/r_{200}$, $c$
is the concentration parameter defined as $c=r_{200}/r_s$,
and $f(c)=c^3/3[\ln(1+c)-c/(1+c)]$. Here $r_{200}$ is the
virial radius within which the average
mass density is 200 times the mean background density (not the critical
density as used in some other studies), and $r_s$ is the scale radius.

The radii of the halos are divided in uniform logarithmic steps
from $0.001r_{200}$ to $2 r_{200}$. 
The number of particles and the average density in each bin are
then computed. Note that when fitting the NFW profiles, we use
only data between a minimum radius, $r_{\rm min}$, and
$r_{200}$. For runs A1, A2 and A4, the minimum
fitting radius $r_{\rm min}$ is $0.02r_{200}$ (corresponding to a
physical radius from 10.5 to 45 \kpc); for run A2 we use
$r_{\rm min}=0.04 r_{200}$ to match the larger force softening length.
The $r_{\rm min}$ value is chosen correspondingly for the PDM simulation.
The best-fit parameters (including the concentration
parameter $c$) are found by minimizing
\begin{equation}
\sum_{i}\left[\log \rho(r_i)- \log\rho_{\rm NFW}(r_i)\right]^2\,.
\end{equation}
Notice that  when we fit the DM profiles in simulations with gas, we scaled the
overdensity $\Delta$ by the DM mass fraction (0.84).

\section{THE INFLUENCE OF NONRADIATIVE GAS ON DENSITY PROFILES}

We choose a large number of massive halos (about 400) to statistically
quantify the systematic influence of the hot gas on the matter
distribution of halos. We only use those halos more massive than
$10^{13}$ {\Msun} (corresponding to $\sim 1.67 \times 10^4$ particles in
the PDM simulation and $3.11\times 10^4$ particles in run A4).

\begin{figure}[b]
\plotone{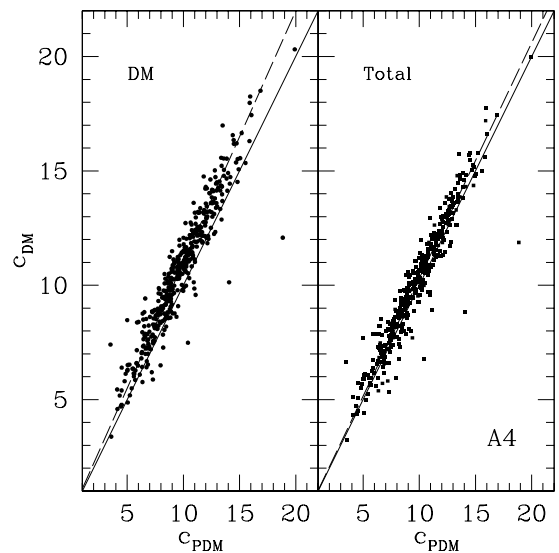}
\caption{Halo concentration parameters in the highest resolution
run, A4, compared with those in the PDM simulation.
In each panel, the filled circles and squares are for the DM
and total mass, respectively. The solid lines indicate equality, while the
dashed lines indicate the concentration parameters that are 10\% ({\em left}) and 3\% ({\em right}) higher than those in the PDM case.
\label{fig1}
}
\end{figure}

The concentration parameters for these halos in our highest resolution
run, A4, versus those in the PDM simulation
are shown in Figure \ref{fig1}. The left and right panels show the concentration
parameters for the DM and total mass distributions,
respectively. For the total mass distribution, 
the concentration parameter is increased only slightly (by about
3\%), while the increase for the DM distribution is
larger, at about 10\%. The effects of baryons (without radiative
cooling)-on the density profile appear to be modest.

To examine how the numerical setups affect the DM and
total mass distributions, in Figure 2 we show the ratios of the
concentration parameters in nonradiative gas simulations relative to
the PDM simulation. 
The median values of this ratio are plotted in five mass bins with equal numbers of halos,
together with the vertical error bars
showing the lower ($25\%$) and upper ($75\%$) quartiles in each bin.
The ratios of the halo concentration parameters in
runs A1 and A2 with those in the PDM run appear to show a
systematic trend with mass for halos with $M\la 3\times
10^{13}h^{-1} M_\odot$. A1 and A2 both have $256^3$ DM particles, a factor of
8 lower than A3 and A4. The inner halo profiles are less
well resolved in A1 and A2. Thus, this poorer resolution results in
the mass dependence and the higher median values of the ratio of the
concentration parameter.
The median values for the ratio of the DM in A1 and A2 are comparable with each other. 
For the A1 run, as a small softening length (relative to the mean
inter-particle separation) was used, one can imagine that 
artificially large accelerations at close encounters may suppress 
structure formation at high redshift in this case, but apparently
the effect has not accumulated significantly to the present epoch.

\begin{figure*}[htpb]
\plotone{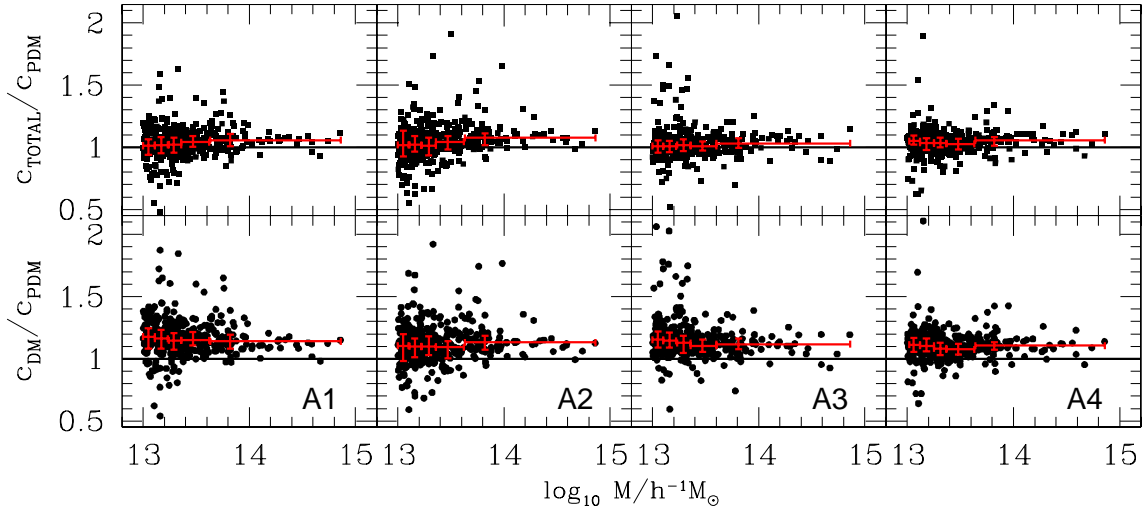}
\caption{Mass dependence of the ratios of halo concentration
parameters in runs A1 -A4 relative to those in the PDM run. The top and
bottom panels are results for the DM and total mass, respectively.
The symbols are the same as Fig.\ref{fig1}. The solid horizontal lines
indicate equality.
The median values and lower and upper quartiles in each bin are indicated
for five mass bins with equal numbers of galaxies. The median value and
the range of mass in each bin are shown as horizontal error bars.
The number of halos is 399, 400, 395, and 410, for the A1, A2, A3, and A4 runs, respectively.
\label{fig2}
}
\end{figure*}

A3 and A4 have the same number of DM particles and force resolution;
they only differ in the number of gas particles ($256^3$ vs. $512^3$).
The third and fourth columns in Figure \ref{fig2} show that A3 and
A4 are in good agreement with each other. It appears to make
little difference for the density profiles of halos whether an equal mass is used for DM and gas particles in simulations.
Furthermore, for both the DM and total mass the ratios of their
concentration parameters to those in the PDM simulation do not
have significant dependence on the halo mass (especially for A4).
The very weak mass dependence and larger scatter in run A3 may be due to the low mass resolution of the gas component.
The consistency between A3 and A4 shows that the resolutions of A4 are sufficient for 
examining the influence of baryons on the density profiles for massive
halos that are well resolved by a large number of particles.

The results we have shown so far are statistical; below we use two
individual halos to see the changes in the halo profiles.
Figure \ref{fig3} shows the density profiles,  $r^2 \rho(r)$, for two
halos with mass of $7.08\times 10^{14}M_\odot$ and $4.16 \times 10^{13}h^{-1} M_\odot$.
The first halo is the most massive one in our simulation. It has a
virial radius of $2.2$\Mpc\,
and contains 1.3 million DM and 1.2 million gas particles in run A4, so it is well resolved.
Figure \ref{fig3} shows that in run A4, the DM and total mass distributions are slightly
steeper than the DM distribution in the PDM run, between $0.02r_{200}$
and $0.12r_{200}$, making the concentration parameter larger.
As expected, the distribution of gas is substantially more extended due
to pressure that resists the gravitational collapse. This is also
reflected in the baryon fraction as a function of radius: in the
central region, the gas mass fraction is only about 10\% at $R\sim 0.02
r_{200}$ (and even smaller at smaller radii) while it
approaches the universal baryon fraction close to the virial radius\citep[see also][]{Ascasibar03}. 
For the group-sized halo shown in the bottom panel of Figure \ref{fig3}
(other low-mass halos show similar behavior), all the simulations have roughly the
same gas density profiles when $R \ga 0.1r_{200}$, but there appears to be a weak
trend for A4 to have a slightly lower gas density at intermediate radius
($r \sim 0.2 r_{200}$) than A1
and A3. However, in the inner regions, the gas density distributions differ
markedly: the highest resolution run, A4, has the highest gas
density, while the lowest resolution run, A1, has the lowest values.

In Figure \ref{fig4}, we plot the temperature and entropy profiles for the two 
halos shown in Figure \ref{fig3}. 
The trend of the temperature as a function of radius is roughly the reverse
of the trend for the density, i.e., A4 has the lowest temperature
in the inner regions.
This is because the gas pressure ($P \propto \rho T$) is well constrained by hydrostatic equilibrium 
(see Fig. 16 in Frenk et al. 1999).
As one can see in the top panels, the temperature profiles
exhibit a flat core toward the halo center, and drop off sharply toward the outer part of halo.
These behaviors are similar to those found by previous studies
\citep{Loken02, Ascasibar03, Rasia04} in spite of the higher mass and force resolutions in our simulations. 
Comparisons between different simulations also reveal that as the resolution increases,
the central temperature decreases while the temperature at large radii  ($r > 0.25 r_{200}$)
remains roughly the same. Our predicted temperature profiles are
somewhat different from the observed ones
(\citep[e.g.,][] {Vikhlinin05}) which often show a similar behavior in the
outer region but a decrease in
the innermost region. The difference likely arises due to radiative cooling, which is not
accounted for in our nonradiative simulations.

In the bottom panels of Figure \ref{fig4}, we plot the profiles of
entropy defined as $kT/n^{2/3}$, where $n$ is the number density of
particles. These entropy profiles are consistent with previous results
\citep[e.g.,][]{Frenk99,Ascasibar03,Kay04,Borgani04,Voit05}.
In particular, the entropy floor in the central region agrees well with
that found by\citet{Ascasibar03} who also performed their simulations with GADGET2 \citep[see also][]{Voit05}.
In the outer region of halos, different SPH simulations yield
consistent results. However, in the inner regions, we find that the
higher the resolution, the lower the entropy floor.
This can be understood as follows: at constant pressure, the entropy $s \propto
T^{5/3}$; in the simulations with lower resolution 
the temperature is higher; as a result, the entropy will also be higher 
in low resolution simulations.
Note results with the same SPH resolution but different DM
resolutions (i.e., run A1 and A3) for the massive halo shown in the
bottom left panel, but for the low-mass halo, the temperature and
entropy in the very inner region in run A1 are somewhat higher than those in run A3.

Two-body heating is expected to be serious for poorly resolved halos
with few particles: gas particles gain energy from collisions
with DM particles and the gas distribution expands substantially. The
trend of differences in the gas density profiles in low-mass halos for
simulations with different resolutions is in
agreement with this two-body heating effect. \citet{Steinmetz97} gave
a rough estimate for when this effect is important (see their eqs. [5] and [6]). 
Even for the low mass halos shown in Fig. \ref{fig3}, the total number of
DM particles ($N$) exceeds 9000 in A1 and 76,000 in A4. Taking $\ln \Lambda = 5$ 
in their equation (6), the two-body heating
timescale is roughly $200 N/10^4$  times the crossing time at half-mass radius,
which may be too long (see also the left panel in their Fig. 2, which shows negligible
two-body heating for a galaxy-sized halo with 4000 particles). 
At least for the halos with mass larger than $10^{13}${\Msun} in our
simulations with highest mass resolution (run A4), two-body heating should be negligible.

\section{ENERGY TRANSFER FROM DM TO GAS PARTICLES IN MERGERS}

It is not entirely clear what causes the steepening in the DM
profiles.
One possibility is that during the assembly process, shocks
produced during mergers between the progenitors can convert kinetic
energy into thermal energy of gas particles. To conserve the total energy, DM
particles lose some of their energy and sink
further toward the center, which steepens the DM density profiles. 
As the merging histories of clusters in cosmological simulations 
are complicated, they are not ideal for isolating the physical mechanism.
Instead, we use nonradiative mergers 
between two spherically symmetric clusters with NFW density profiles
to demonstrate the energy exchange between DM and gas particles.

We utilize the simulations of \citet{McCarthy06}. These simulations are gas-only and gas+DM simulations of head-on mergers with two equal-mass clusters using GADGET2. 
In the latter case, the gas traces DM initially and the ratio of gas to 
DM mass is set to $\Omega_b/\Omega_m\approx 16\%$. The virial mass of each cluster is
$10^{15}M_{\odot}$ within $r_{200}$. 
In these simulations each cluster has $50,000$ gas particles
and $50,000$ DM particles when present.
The two clusters are initially well separated, with a distance much larger than $r_{200}$.
The mergers start with the same infall velocity and separation
in the gas-only and gas+DM simulations, so the initial masses and total energy are the same. 
The force-softening length is set to $10$\,kpc for both the gas and DM particles.
More details, including the setup of initial conditions, can be
found in \citet{McCarthy06}.

Figure \ref{fig5} shows the resulting binned gas 
density profiles from the gas-only and gas + DM simulations 
after evolution for a Hubble time.  
Both the results of the gas-only and gas + DM simulations are
similar outside $r \sim 150$\,kpc. But inside this radius,
the gas density is higher in the gas + dark matter simulation, implying that 
the dark matter must have transferred energy to the gas inside this radius.
Figure \ref{fig6} shows the
evolution of the total energy of the DM from the 
gas + DM simulation. The energy has been normalized by its 
initial value.  Note that the total energy is negative (the system is 
gravitationally bound), so if the ratio exceeds unity this implies that
energy has been transferred to the gas.  
The figure clearly demonstrates that at the end of this simulation, 
approximately 7\% of the DM's total initial energy has been 
transferred to the gas. The same energy transfer procedure should
have also occurred in the cosmological simulations.

\section{SUMMARY AND DISCUSSION}

In this paper, we performed a set of numerical simulations with
nonradiative gas in the concordance cosmology, and compared the results with those of a pure DM
simulation. The simulations have identical initial conditions but
different force and mass resolutions. We carefully examined the
effects of force resolutions and two-body heating. In summary, in
simulations with hot gas, the DM distribution
is more concentrated than those in DM-only simulation.
The DM and total mass concentration parameters in the nonradiative gas simulations
are on average about 10\% and $\sim 3\%$ larger than those in the PDM simulation.
We used idealized simulations of mergers of two clusters to demonstrate
that the steepening of the DM profile is due to 
energy transfer from DM to gas particles (in shocks), 
which should also occur in our cosmological simulations.

The influence of baryons on the total mass distribution in our simulations is less
significant than the results by \cite{Rasia04} who reported
a 10\% increase in the concentration parameter for the total mass
distribution.
The difference is likely due to different SPH treatments
used; in GADGET2 the entropy is conserved, while in GADGET1 it is not.
New results obtained using GADGET2 by the same group
are consistent with ours (A. Rasia 2006, private communication).
Note also that we have many halos (about 400) while they had only 17 halos,
so our statistics are better.
Moreover, \citet{Ascasibar03} also showed (their Fig. 1) that gas
densities at the inner part of halos are higher in a
conventional SPH code than those obtained in an entropy-conserving SPH code.
Overall, the influence of hot gas on the mass distribution
is quite weak in nonradiative gas simulations, especially for the total mass.
Observational measurement of this effect will be challenging. For example,
the small change in the total mass concentration parameters 
will not have a substantial impact on the cross section of giant arcs.

\begin{figure}
\begin{center}
{
\includegraphics[width=.5\textwidth]{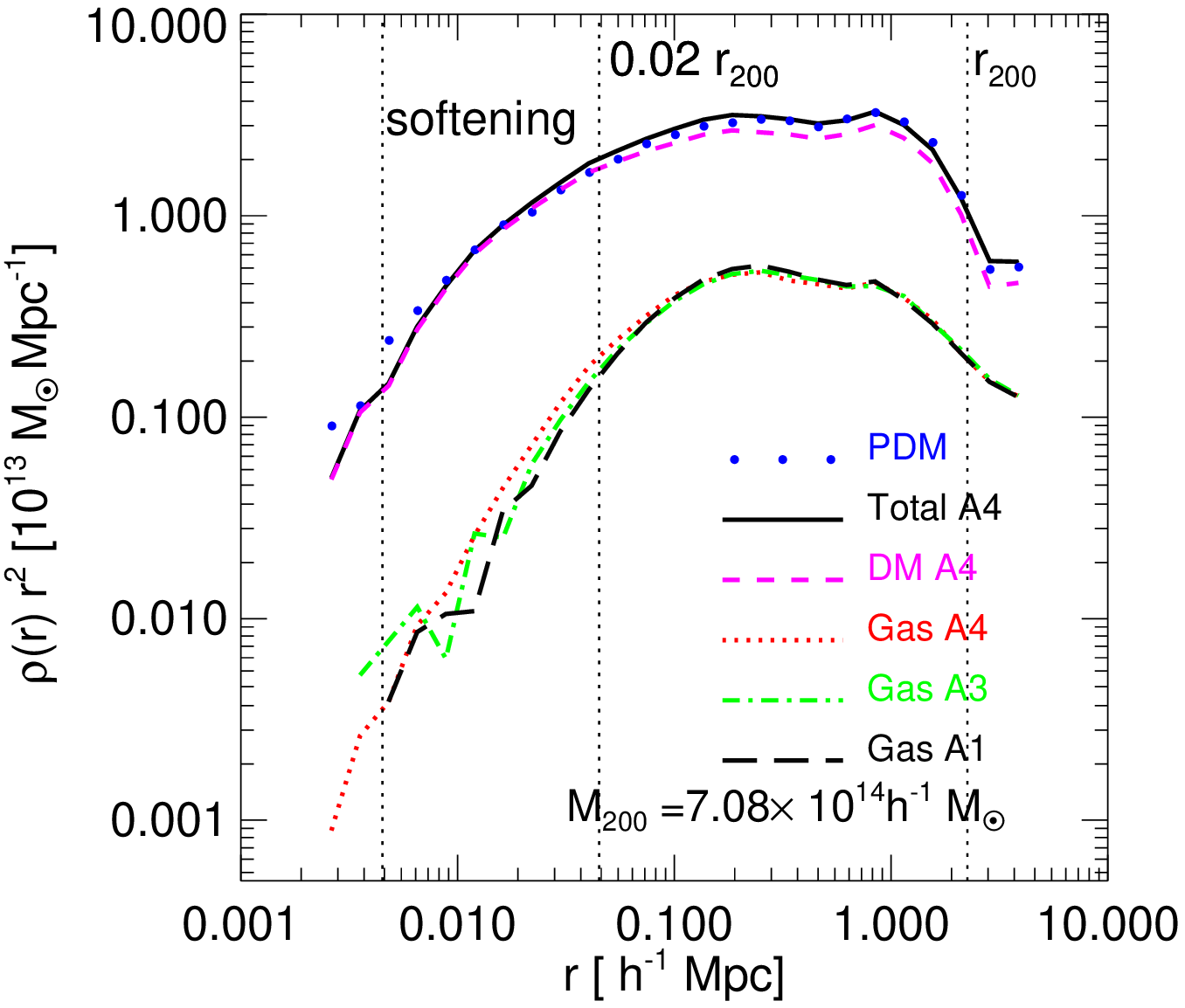}
\\[5mm]
\includegraphics[width=.5\textwidth]{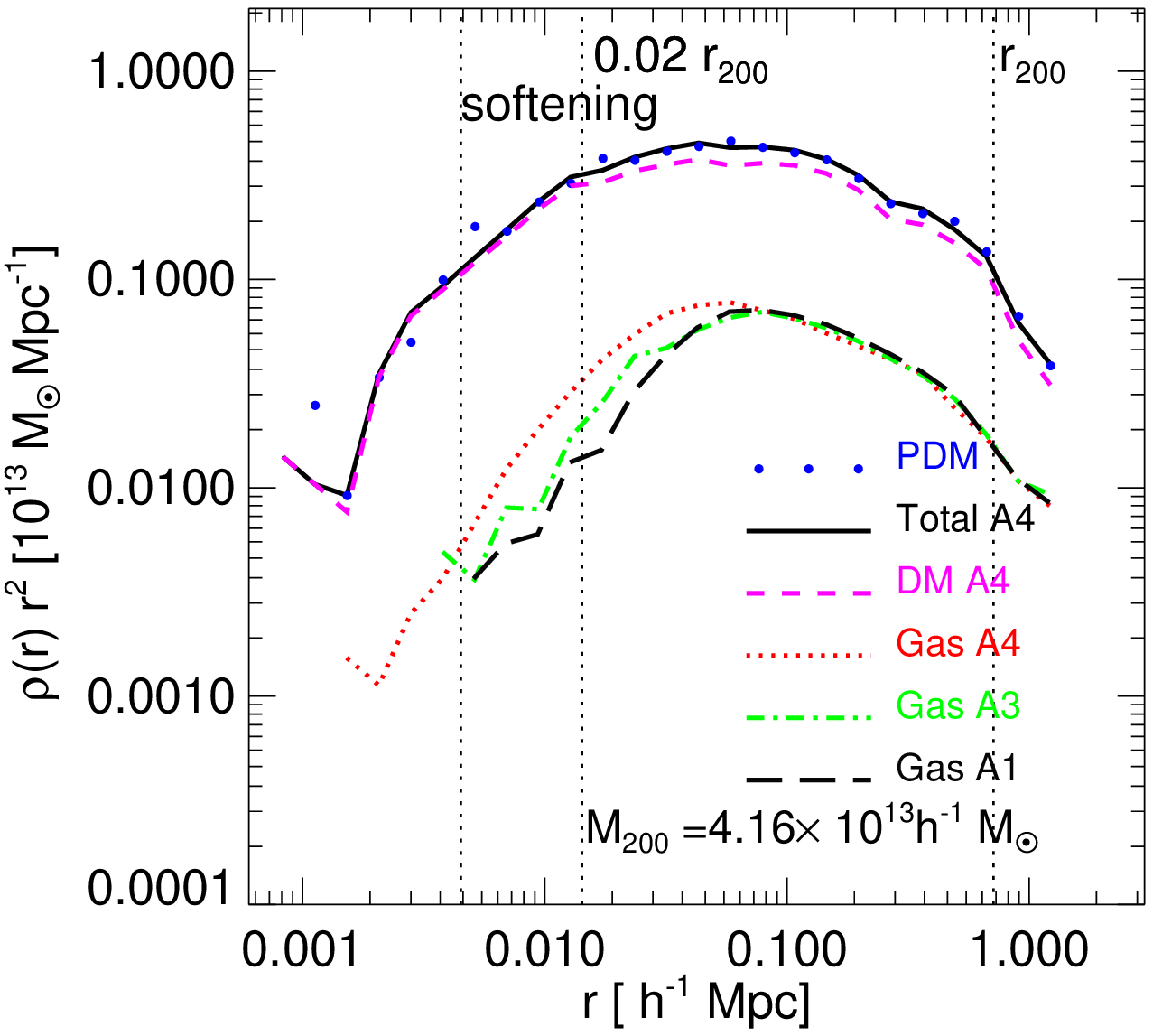}
}
\end{center}
\caption{Density profiles for the most massive halo ({\em top}) and a group-sized halo ({\em bottom}) in our simulations.
The filled dots indicate the density profile for the PDM run, while
other lines indicate results for the total, DM and gas density profiles
in runs A1, A3, and A4 (labeled inside the panels). The three dotted
vertical lines indicate the softening length, $0.02 r_{200}$ and
$r_{200}$. The data points between the latter two are used in the profile
fitting in A1, A3, and A4.
\label{fig3}
}
\end{figure}

\begin{figure}
\begin{center}
{
\includegraphics[width=.5\textwidth]{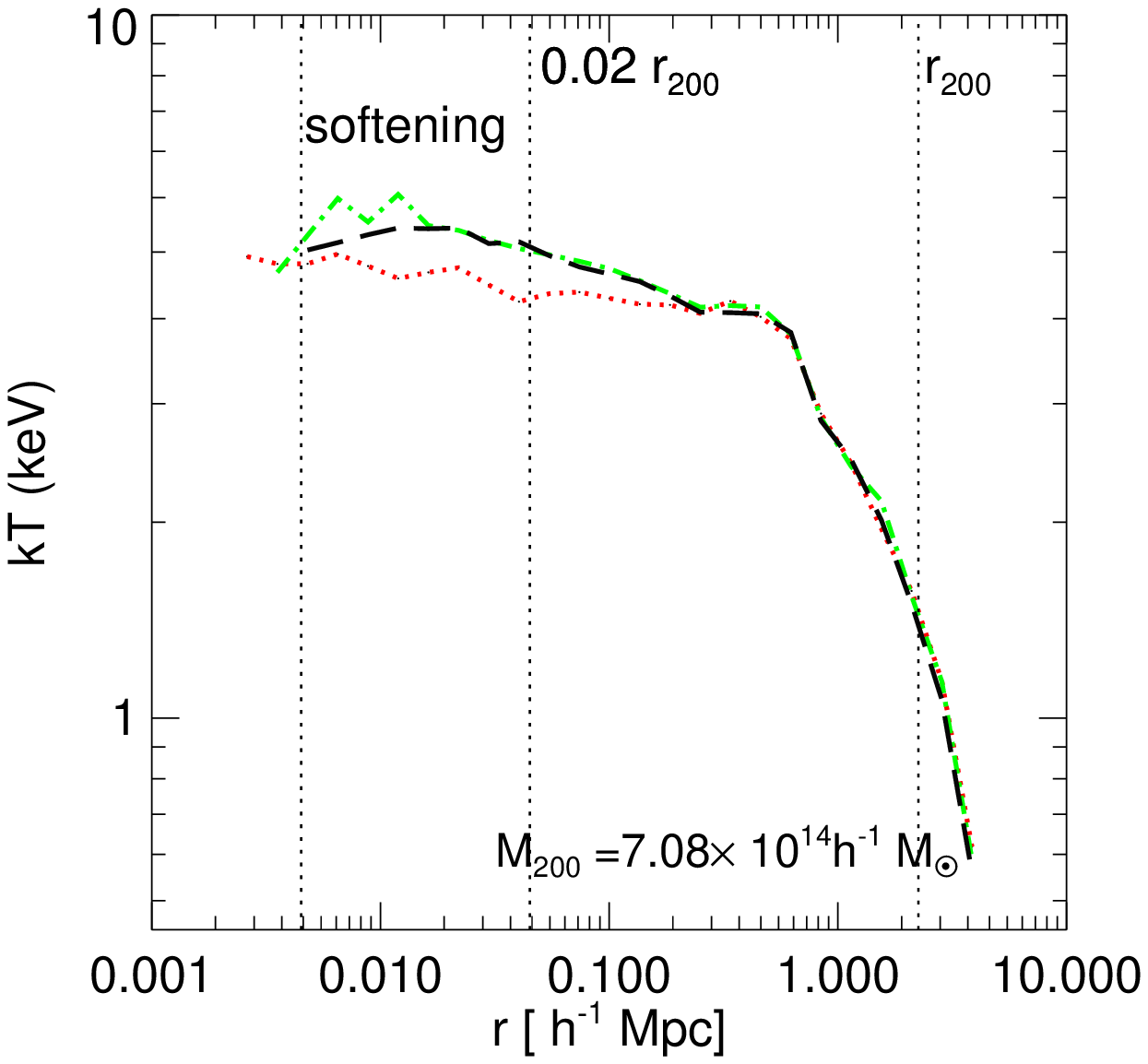}\includegraphics[width=.5\textwidth]{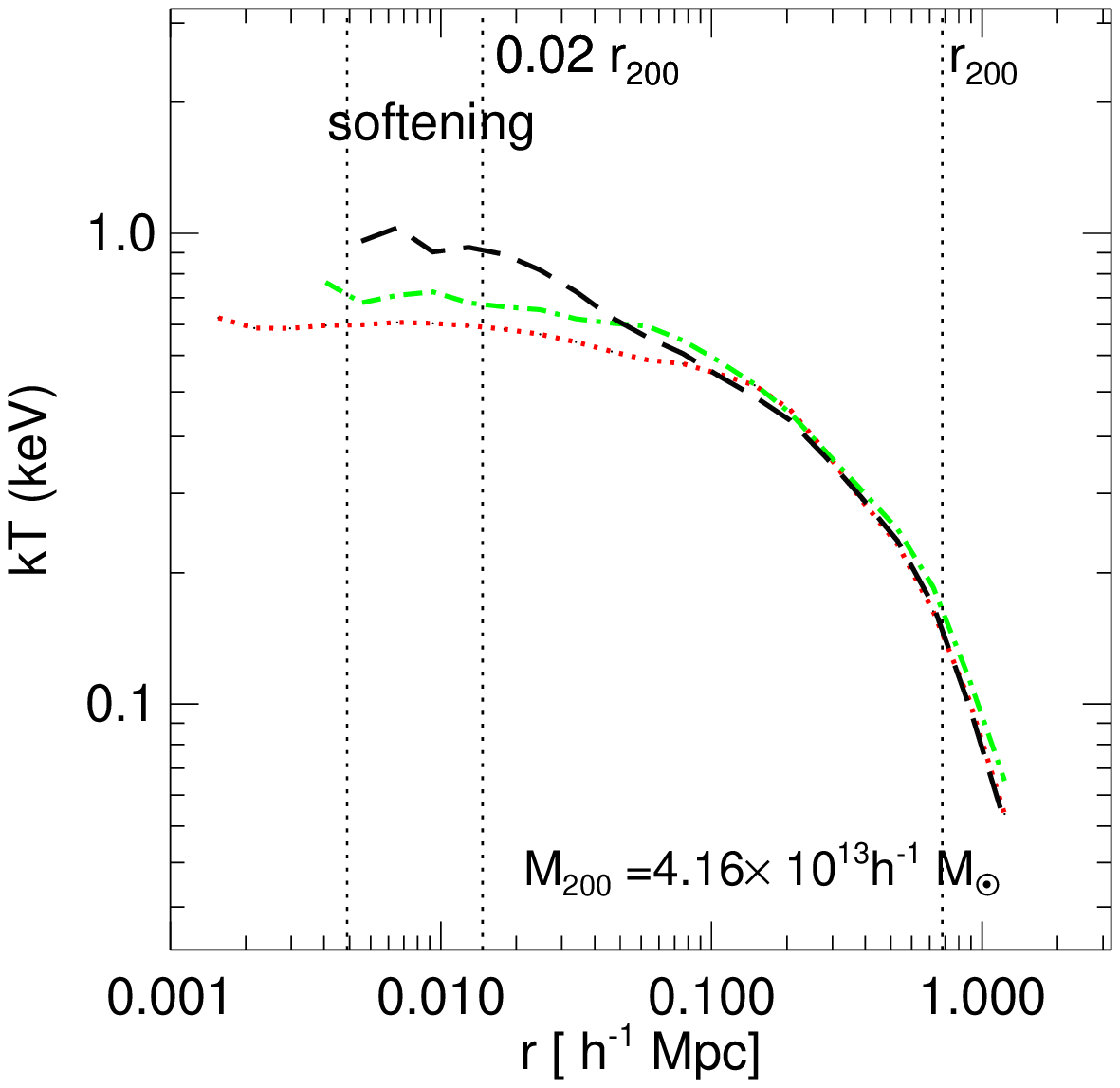}
\\[5mm]
\includegraphics[width=.5\textwidth]{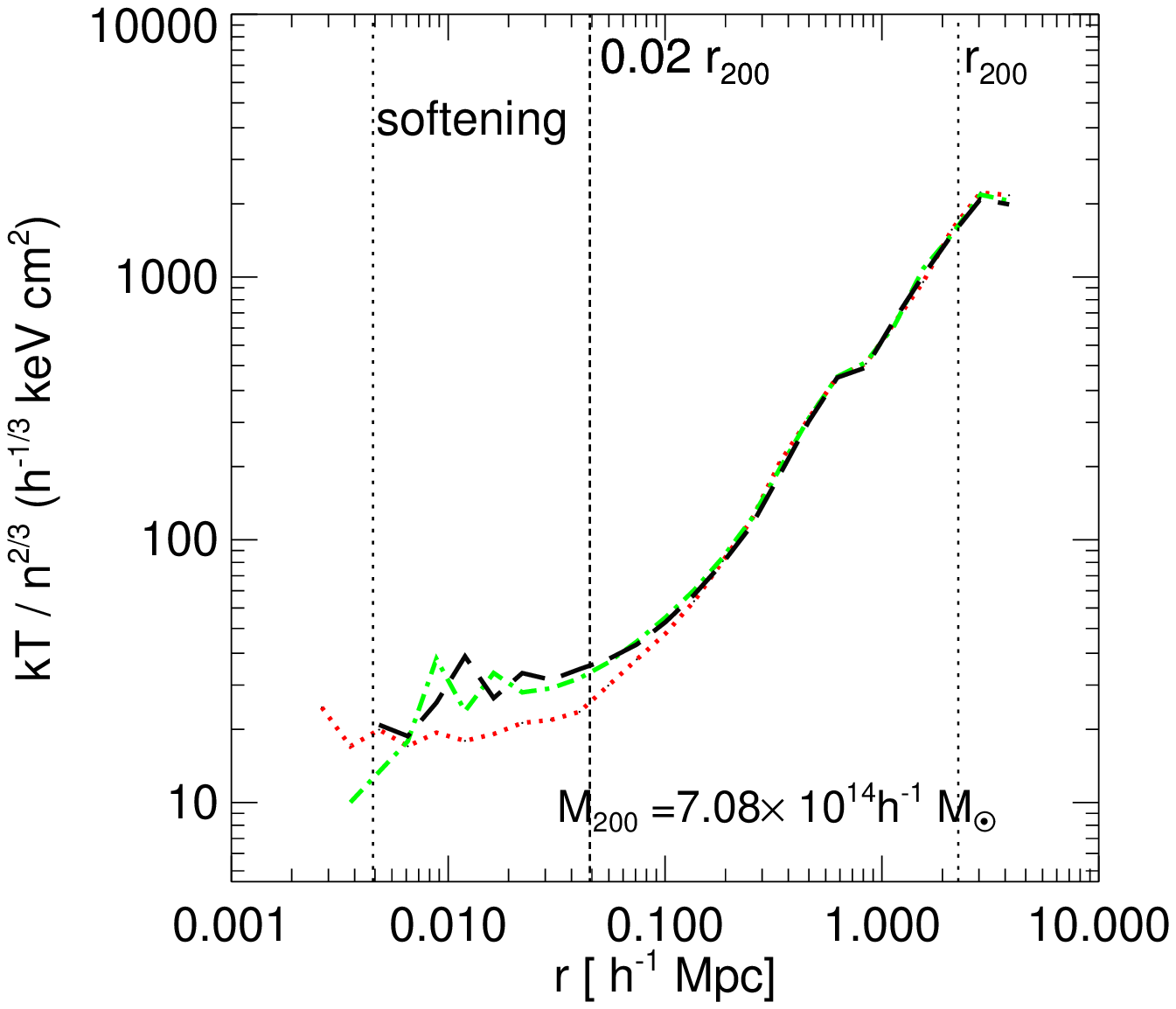}\includegraphics[width=.5\textwidth]{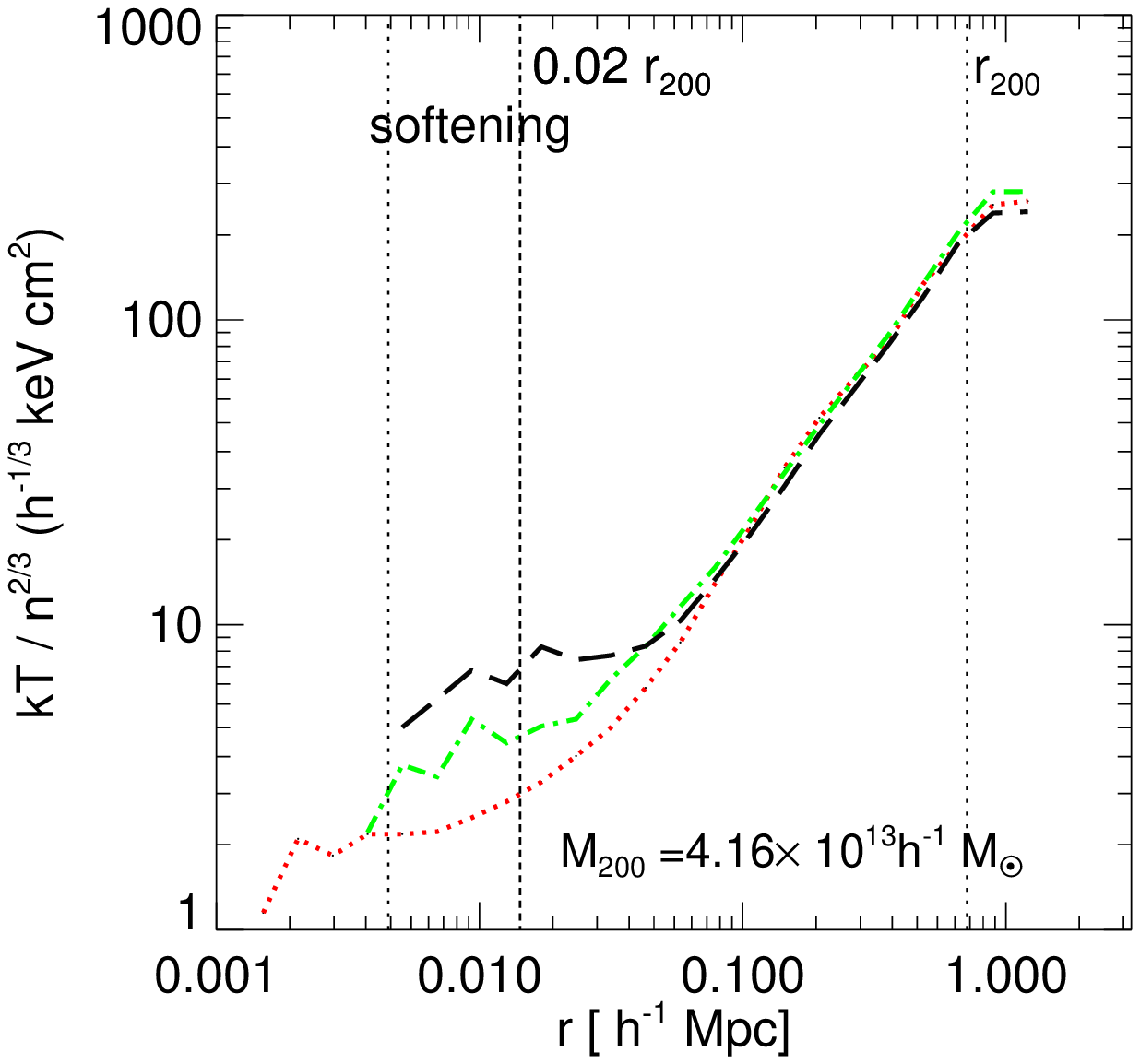}
}
\end{center}
\caption{
Temperature ({\em top}) and entropy profiles ({\em bottom}) for the two halos shown in Fig. \ref{fig3}. The line symbols are the same as in Fig. \ref{fig3}.
\label{fig4}
}
\end{figure}

\begin{figure}
\begin{center}
\includegraphics[width=0.58\textwidth]{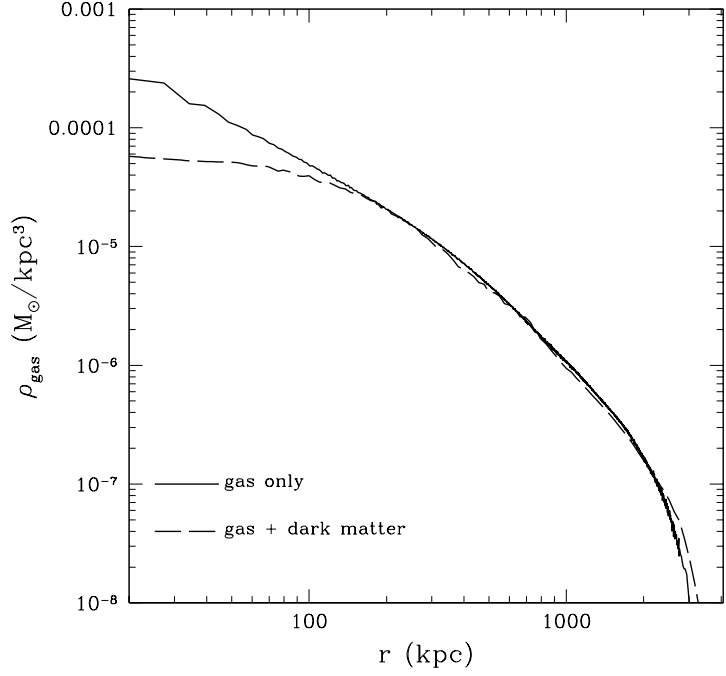}
\end{center}
\caption{
Gas density profiles in the gas-only ({\em solid line}) and in the gas
+ DM  ({\em dashed line}) simulations of
head-on collisions between two spherical
NFW clusters (see \S 4 for more details).
\label{fig5}
}
\end{figure}

\begin{figure}
\begin{center}
\includegraphics[width=0.58\textwidth]{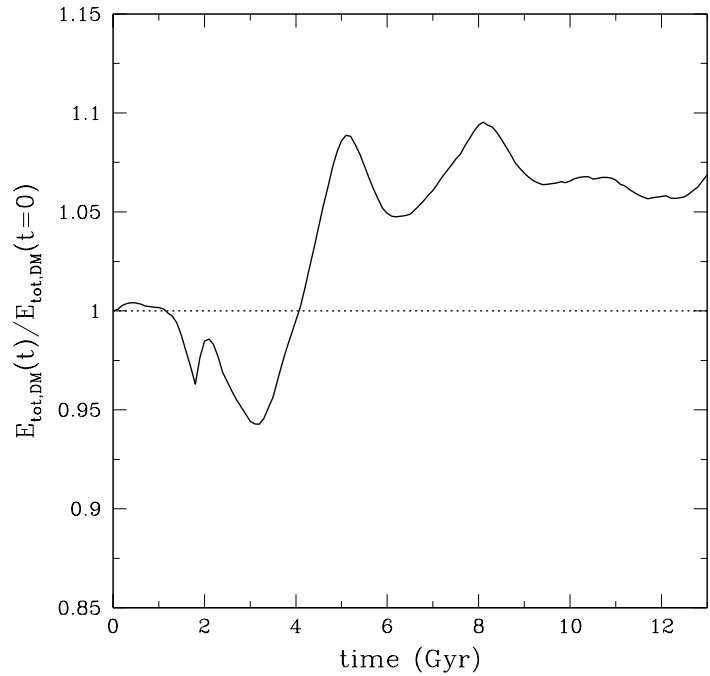}
\end{center}
\caption{
Ratio of the total energy of the DM at time $t$ to the initial
total energy in the gas + DM simulation.
The dotted line indicates energy conservation.
At $t\ga 5$ Gyr, the dark matter has lost energy to the gas, as implied by
a ratio larger than unity.
\label{fig6}
}
\end{figure}

The gas density profiles in our simulations show considerable
scatter in their inner parts in different simulations, well beyond the softening length
(see Fig. \ref{fig3}). 
We also show the results of temperature and entropy profiles to explain the effects of 
the resolution of SPH particles.
While two-body heating of gas particles by DM
particles is qualitatively consistent with the trend we see,
quantitatively the timescale may be too long for two-body heating to have substantial
effects on the gas density profiles, in particular for the halos studied in run A4. 
Notice, however, that the uncertainty in gas profile has little effect on the DM,
and total mass distributions in our results as gas only contributes a
small (but somewhat uncertain) fraction of the total mass.

The simulations presented above neglected important
physical processes such as star formation, supernovae feedback, and heat
conduction. For the  group- and cluster-scale halos that we analyzed, the
processes mentioned above are expected to be less important than in
galaxies, nevertheless, their proper treatment is important in
comparing data and theoretical predictions.
Our study also ignores the differences in dynamical frictions between
galaxies and DM halos. Because galaxies are located at the centers of DM halos, they
can survive the tidal disruptions longer than the DM halos
(e.g., \citealt{Gao04}).
As the dense cores of galaxies sink toward the centers, their kinetic
energy can be transferred to and heat up the DM particles to make the DM distribution flatter \citep{Zant01,Zant04}.
The uncertainties in the gas density profiles we found in the nonradiative
case serves as a caution that while the incorporation of star formation and
feedback processes is desirable, their physical treatment is likely to
add another layer of uncertainty.

\acknowledgments

We acknowledge the anonymous referee for constructive comments which 
improved the paper.
We thank Volker Springel for useful comments and suggestions on the draft 
as well as kindly providing the code GADGET2 before public release.
W.P.L. thanks Kohji Yoshikawa and Mei Zhang for help on the simulations.
This work is supported by grants from NSFC (10203004, 10125314,
10373012, and 10533030) and Shanghai Key Projects in Basic Research
(04JC14079 and 05XD14019). S.M. and L.G. acknowledge travel support from
CAS and Jodrell Bank. 
I.G.M. acknowledges support from an NSERC postdoctoral fellowship.
The simulations were performed at the Shanghai Supercomputer Center.

\clearpage

\clearpage

\clearpage

\begin{thebibliography}{99}

\bibitem[Ascasibar et al. (2003)]{Ascasibar03} Ascasibar, Y., Yepes, G., M\"uller, V., \& Gottl\"ober, S., 2003, \mnras, 346, 731

\bibitem[Bartelmann \& Meneghetti (2004)]{Bartelmann04} Bartelmann, M., \& Meneghetti, M. 2004, \aap, 418, 413

\bibitem[Blumenthal et al. (1986)]{Blumenthal86} Blumenthal, G.R., Faber S.M., Flores, R., \& Primack, J.R. 1986, \apj, 301, 27

\bibitem[Borgani et al. (2004)]{Borgani04} Borgani, S., et al. 2004, \mnras, 348, 1078

\bibitem[Buote \& Lewis (2004)]{Buote04} Buote, D.A., \& Lewis, A.D., 2004, \apj, 604, 116

\bibitem[Broadhurst et al.(2005)]{Broadhurst05} Broadhurst, T., Takada, M., Umetsu, K., Kong, X., Arimoto, N., Chiba, M., Futamase, T. 2005, \apjl, 619, L143

\bibitem[Dalal \& Keeton (2003)]{Dalal03} Dalal, N., \& Keeton, C.R., preprint (astro-ph/0312072)

\bibitem[Dalal et al. (2004)]{Dalal04} Dalal, N., Holder, G., \& Hennawi, J.F. 2004, \apj, 609, 50

\bibitem[El-Zant et al. (2004)]{Zant04} El-Zant, A., Hoffman, Y., Primack, J., Combes, F., \& Shlosman, I. 2004, \apjl, 607, L75

\bibitem[El-Zant et al. (2001)]{Zant01} El-Zant, A., Shlosman, I., \& Hoffman, Y., 2001, \apj, 560, 336

\bibitem[Frenk et al. (1999)]{Frenk99} Frenk, C.S. et al. 1999, \apj, 525, 554
 
\bibitem[Gao et al. (2004)]{Gao04} Gao, L., White, S.D.M., Jenkins, A., Stoehr, F.,  Springel, V. 2004, \mnras, 355, 819

\bibitem[Gnedin et al.(2004)]{Gnedin04} Gnedin, O.Y., Kravtsov, A.V., Klypin, A.A., Nagai, D. 2004, \apj, 616, 16

\bibitem[Hennawi et al. (2006)]{Hen06} Hennawi, J. F., Dalal, N., Bode,
	P., Ostriker, J. P. 2006, in press (astro-ph/0506171)

\bibitem[Horesh et al. (2005)]{Horesh05} Horesh, A., Ofek, E. O., Maoz,
	D., Bartelmann, M., Meneghetti, M.,  Rix, H.-W. 2005, \apj, 633, 768

\bibitem[Jing \& Suto (2000)]{JS00} Jing, Y.P., \& Suto, Y. 2000, \apjl, 529, L69

\bibitem[Jing et al. (2006)]{Jing06} Jing, Y.P., Zhang, P.J., Lin, W.P., Gao, L., \& Springel, V. 2006, \apjl, 640, L119

\bibitem[Kay (2004)]{Kay04} Kay, S.T. 2004, \mnras, 347, L13

\bibitem[Li et al. (2005)]{Li05} Li, G.L., Mao, S., Jing, Y. P., Bartelmann, M., Kang, X., Meneghetti, M. 2005, ApJ, 635, 795

\bibitem[Lewis et al. (2003)]{Lewis03} Lewis, A.D., Buote, D.A., \& Stocke, J.T., 2003, \apj, 586, 135

\bibitem[Loken et al. (2002)]{Loken02} Loken, C., et al. 2002, \apj, 579, 571

\bibitem[McCarthy et al. (2006)]{McCarthy06} McCarthy, I.G., et al. 2006, in preparation

\bibitem[Meneghetti et al. (2005)]{Meneghetti05} Meneghetti, M., Bartelmann, M., Jenkins, A., \& Frenk, C. 2005, preprint (astro-ph/0509323)

\bibitem[Mo et al. (1998)]{Mo98} Mo, H., Mao, S., \& White,
	S.D.M. 1998, \mnras, 295, 319

\bibitem[Moore et al. (1998)]{Moore98} Moore, B., Governato, F., Quinn, T., Stadel, J., \& Lake, G. 1998, \apjl, 499, L5

\bibitem[Navarro et al.(1995)]{Navarro95} Navarro, J.F., Frenk, C.S., \& White, S.D.M. 1995, \mnras, 275, 720

\bibitem[Navarro et al. (1997)]{NFW97} Navarro, J.F., Frenk, C.S., \& White, S.D.M. 1997, \apj, 490, 493

\bibitem[Pearce et al. (1994)]{Pearce94} Pearce, F.R., Thomas, P.A., \& Couchman 1994, \mnras, 268, 953

\bibitem[Pearce et al. (2000)]{Pearce00} Pearce, F.R., Thomas, P.A., Couchman, H.M.P.,\& Edge, A.C. 2000, \mnras, 317, 1029

\bibitem[Pointecouteau et al.(2005)]{Pointecouteau05} Pointecouteau, E., Arnaud, M. \& Pratt,G.W. 2005, \aap, 435, 1

\bibitem[Rasia et al.(2004)]{Rasia04} Rasia, E., Tormen, G., Moscardini L. 2004, \mnras, 351, 237

\bibitem[Sand et al. (2004)]{Sand04} Sand, D.J., Treu, T., Smith, G.P., \& Ellis, R.S. 2004, \apj, 604, 88

\bibitem[Spergel et al. (2003)]{Spergel2003} Spergel, D.N., et al. 2003, \apjs, 148, 175

\bibitem[Spergel et al. (2006)]{Spergel2006} Spergel, D.N., et al. 2006, preprint (astro-ph/0603449)

\bibitem[Springel (2005)]{Springel05} Springel, V. 2005, MNRAS, 364, 1105

\bibitem[Springel \& Hernquist (2002)]{Springel02} Springel, V., \& Hernquist, L. 2002, \mnras, 333, 649 

\bibitem[Springel et al. (2001)]{Springel01} Springel, V., Yoshida, N., \& White, S.D.M. 2001, New Astronomy, 6, 79

\bibitem[Steinmetz \& White (1997)]{Steinmetz97} Steinmetz, M., \& White, S.D.M. 1997, \mnras, 288, 545

\bibitem[Tyson et al. (1998)]{Tyson98} Tyson, J.A., Kochanski, G.P., \& dell\'Antonio, I.P. 1998, \apjl, 498, L107

\bibitem[Vikhlinin et al. (2005)]{Vikhlinin05} Vikhlinin, A., et al. 2005, \apj, 628, 655

\bibitem[Voit et al.(2005)]{Voit05} Voit, G.M., Kay, S.T., \& Bryan, G.L., 2005, \mnras, 364, 909

\bibitem[Zhan \& Knox (2004)]{Zhan04} Zhan, H. \& Knox, L. 2004, \apjl, 616, L75

\end{thebibliography}
\end{document}